# The Range Shrinks, the Threat Remains: Re-evaluating LLM Package Hallucinations on the 2026 Frontier-Model Cohort

Aleksandr Churilov · Independent Researcher · churik0509@gmail.com

**Abstract**

Package hallucinations, LLM-generated references to packages that do not exist, enable slopsquatting: an attacker registers the hallucinated name. We replicate Spracklen et al. (USENIX Security ’25) on five 2026 frontier models over 199,840 Python and JavaScript generations. This version reports a preregistered live revalidation that replaces the static package lists of earlier versions with two independent checks against the live PyPI and npm registries. Hallucination rates are 3.6–5.0% of package mentions (2.2–3.4% of generations): the 2026 frontier sits at the level of 2024’s commercial models, and the open-weight gap seen in 2024 has disappeared. Python exceeds JavaScript for every model. The set of names hallucinated by all five models shrinks from the 127 reported earlier to 104; 26 earlier entries were correct import statements whose module name differs from the package name. Using the accidental six-month public exposure of our own list as a natural experiment, we find no detectable effect of publication: about 1% of hallucinated names were registered within six months, at the same rate for published and unpublished names, and the 95% interval excludes increases above 2.5 percentage points. Still, at least 23 package names that the models suggest had previously hosted malware later removed by the registries, and one hallucinated name was registered with malware after our runs.

*Version 4 (November 2026). This version replaces the validation of earlier versions (static package lists) with a preregistered live revalidation, and it supersedes all numbers in versions 1–3 and in the SCORED ’26 workshop version. Section 7 lists every changed number and the reason for the change.*

## 1 Introduction

Modern software development depends on centralized package registries: PyPI for Python and npm for JavaScript host millions of packages installable with a single command. Code-generating large language models have added a new actor to this supply chain. When a developer asks an LLM for code, the model often emits `pip install` or `npm install` directives, or import statements that name particular packages. If the developer trusts the suggestion and installs the named package, an attacker who has registered a malicious package under that name succeeds without ever interacting with the developer or the LLM provider.

Spracklen et al. [1] characterized this risk on the 2024 model cohort, generating 576,000 code samples across 16 models. Hallucinated package references averaged 5.2% for commercial models and 21.7% for open-source models, and the authors identified 205,474 unique hallucinated package names. The resulting attack is now called *slopsquatting* [18]: registering malicious packages under names that LLMs have, by their hallucination behavior, primed developers to install.

Two model generations later, Anthropic, OpenAI, Google and DeepSeek have released code-capable models with larger training sets, broader safety post-training and, in several cases, reasoning-effort controls. We measure whether package hallucination has changed with them.

We report five findings. First, the 2026 frontier sits at the level of 2024’s best commercial models: per-model rates range from 3.63% to 5.02% of package mentions, between the lowest rate Spracklen et al. report for any model (3.59%, GPT-4 Turbo) and their commercial average (5.2%), and the open-weight model in our cohort is no longer an outlier. Second, 104 package names (88 PyPI, 16 npm) are hallucinated by all five models. Third, Python rates exceed JavaScript rates for every model, the reverse of the 2024 pattern. Fourth, within the Anthropic family, Claude Haiku 4.5 hallucinates less than Claude Sonnet 4.6. Fifth, hallucinated names do get registered, but rarely and not faster when published: about 1% of hallucinated names were registered within six months, at the same rate for names whose list became public by accident and for names that were never published. At least 23 names the models suggest had already hosted malware that the registries removed.

This version also corrects its predecessors. Versions 1–3 validated names against static package lists from 2025 and counted the import names of real packages as package names. A preregistered live revalidation against both registries (§4.5, §4.9) lowers every model's rate by about a fifth and shrinks the universal set from 127 to 104 names; Section 7 reports every change and its cause. The qualitative findings of earlier versions hold.

**Contributions.**

- A replication of Spracklen et al. [1] on five frontier LLMs released between October 2025 and March 2026 (199,840 generations), validated against the live PyPI and npm registries with two independent methods per registry.
- Per-model rates at two levels (package mentions, comparable with [1], and generations, as preregistered), showing that the frontier now clusters at the level of 2024's best commercial models.
- A universal-hallucination set of 104 names that all five models invent, broken down by category, together with the assessment by PyPI Security and Socket of which names an outsider can register.
- A preregistered natural experiment on whether publishing hallucinated names speeds up their registration (it does not, at this precision), and evidence that at least 23 names the models suggest had previously hosted malware.
- An open artifact: code, the preregistration with its deviation log, and all aggregate results (§11.3).

The remainder of the paper is organized as follows. §2 reviews related work; §3 states the threat model; §4 describes the methodology; §5 presents per-model results; §6 examines the universal set; §7 reports what changed since earlier versions; §8 measures registrations of hallucinated names; §9 discusses the findings; §10 lists limitations; §11 covers ethics and open science; §12 concludes.

## 2 Background and Related Work

### 2.1 Package Confusion Attacks

Package confusion attacks predate LLM-generated code by more than a decade. Typosquatting registers packages under names one or two edits away from a legitimate package, exploiting developer typos at install time; Neupane et al. [14] catalog thousands of such attacks on PyPI and npm. Birsan's 2021 dependency-confusion disclosure [15] showed that internal package names, shadowed on a public registry, can be silently substituted by the package manager. Slopsquatting [18] is the LLM-induced variant: instead of waiting for a typo or a name leak, the adversary registers names that the LLM itself has primed developers to type. This paper continues the measurement line of [1] on the size of the attack surface that current LLMs create.

### 2.2 Hallucinations in Code-Generating LLMs

LLM hallucination has been studied extensively for natural-language output [16, 17]. Code-generation hallucination is a security-relevant special case: the model emits syntactically plausible but wrong tokens, and when those tokens are package names they become an installation request rather than a textual claim. Pearce et al. [13] established the underlying pattern for vulnerable code generated by GitHub Copilot. Related work [20] extends the package-hallucination measurement to more models. The present paper is a temporal replication: it runs the same prompts on newer models in order to compare them with the 2024 baseline.

### 2.3 The Spracklen Baseline

Spracklen et al. [1] is the baseline for this work. They evaluated 16 LLMs released between mid-2023 and mid-2024 on two prompt datasets, 576,000 generations in total. Hallucination rates averaged 5.2% for commercial and 21.7% for open-source models; the lowest per-model rate was 3.59% (GPT-4 Turbo). They identified 205,474 unique hallucinated names and characterized them by repeatability, similarity to real packages, persistence across temperatures and recurrence across models. Their public artifact [19] contains the two prompt datasets, package-name lists for PyPI and npm, and the extraction and validation code. Earlier versions of this paper validated names against those lists; this version validates against the live registries (§4.5).

### 2.4 Concurrent and Subsequent Work

Importing Phantoms [20] reports comparable rates on a partially overlapping model set. Industry monitoring services have published rate measurements in 2025–2026 [18]. Spracklen, Aghazadeh, Koushanfar and Jadliwala [21] suppress package hallucinations with adaptive unlearning; their method uses the registry as an oracle that labels names as valid or hallucinated, and they note that an attacker who pre-registers hallucinated names inverts that signal ("oracle poisoning"). Section 8 estimates how often that precondition occurs in practice. Spracklen et al. [1] already measured how often the same hallucinated name recurs across models in their cohort and found that 81% of distinct hallucinated names came from a single model; we report the intersection across all five 2026 frontier models and its registrability as assessed by PyPI Security and Socket.

## 3 Threat Model

We adopt the threat model of Spracklen et al. [1]. The adversary is an unprivileged actor whose only required capability is registering packages on PyPI and npm. Anyone with an email address can do this, and neither registry reviews first-time uploads. The adversary's goal is to publish a malicious package under a name that one or more frontier LLMs are likely enough to hallucinate that developers, prompted with similar queries, will install it. The attack chain (Figure 1) has six stages:

1. the adversary samples one or more LLMs and identifies hallucinated package names;
2. the adversary registers a malicious package on PyPI or npm under each such name;
3. later, a victim developer issues a code-generation query to the same or a related LLM;
4. the LLM generates code that references the hallucinated name;
5. the developer's package manager, trusting the registry, installs the adversary's package;
6. the malicious code executes in the developer's environment.

We assume no compromise of the LLM provider, no fine-tuning by the adversary, no privileged access to user systems, and no involvement of the registries beyond their public interfaces. Uploads are free, so the adversary's only cost is the LLM API spend needed to enumerate hallucinations; our pipeline's own cost estimate (Appendix A) puts it at about $1,160 for a sweep of comparable scope.

The universal-hallucination set of §6 strengthens the adversary's position: one registered name can target users of several providers at once, since 104 names in our data are hallucinated by all five models. Whether a name can actually be registered depends on each registry's own rules (§6).

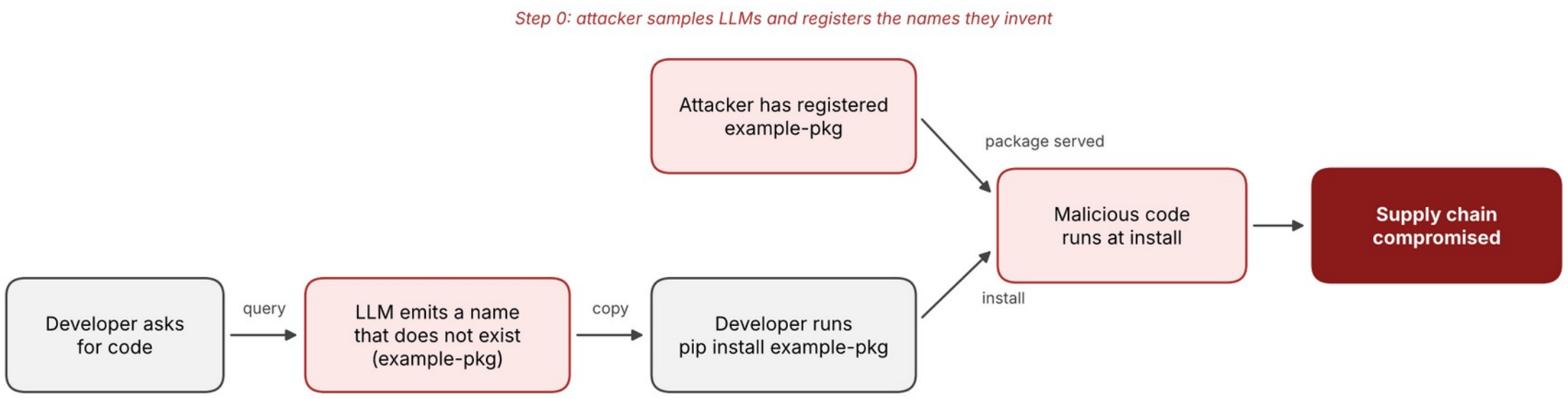

*Figure 1: Slopsquatting attack chain. The developer issues a code-generation query; the LLM emits an install or import directive that names a package that does not exist (shown with a placeholder name); an attacker who has registered that name wins code execution at install time. Adapted from Spracklen et al. [1].*

# 4 Methodology

## 4.1 Models Studied

We evaluate five frontier code-capable LLMs released between October 2025 and March 2026 that represent the major commercial and open-weight providers: Anthropic's Claude Sonnet 4.6 [2] and Claude Haiku 4.5 [3], OpenAI's GPT-5.4-mini [4], Google's Gemini 2.5 Pro [5], and DeepSeek V3.2 [6, 7]: four providers, three closed-weight and one open-weight. Table 1 summarizes the models. Compared with Spracklen's 2024 cohort [1], this is an eighteen-month forward shift and a deliberate restriction to frontier-tier endpoints.

*Table 1: Models studied. Snapshot identifiers are the model strings returned by each provider's API during the experimental run (April 22–28, 2026). Refusal rates are recomputed from the stored generations with the definition of §4.8.*

| Provider | Model | Snapshot ID | Reasoning | Refusals | Spend |
|---|---|---|---|---|---|
| Anthropic | Claude Sonnet 4.6 | claude-sonnet-4-6 | default, no extended thinking | 18.77% | Appx. A |
| Anthropic | Claude Haiku 4.5 | claude-haiku-4-5-20251001 | default, no extended thinking | 1.27% | Appx. A |
| OpenAI | GPT-5.4-mini | gpt-5.4-mini-2026-03-17 | minimal | 31.51% | Appx. A |
| Google | Gemini 2.5 Pro | gemini-2.5-pro | default | 8.11% | Appx. A |
| DeepSeek | DeepSeek V3.2 | deepseek-chat (alias; see §10) | default | 1.97% | Appx. A |

## 4.2 Prompt Datasets

We re-use the two prompt datasets released by Spracklen et al. [1] without modification: 20,163 programming questions drawn from Stack Overflow (subsets SO_AT and SO_LY) and 19,806 LLM-synthesized questions (subsets LLM_AT and LLM_LY), each split evenly between Python and JavaScript. We generate one sample per prompt per model. The stored corpus has 199,840 generations: 39,969 for each model except Claude Sonnet 4.6, for which five generations are missing (Appendix A). We add no new 2026 prompts, so that the corpus matches Spracklen's exactly and the comparison is direct. We discuss the resulting training-data leakage concern in §10.

## 4.3 Generation Configuration

For each model we record the API endpoint, the snapshot identifier returned in the response metadata, the temperature, the top-p value and, where available, the reasoning-effort setting. Defaults were used unless noted. GPT-5.4-mini was run with reasoning_effort set to minimal, a choice made for cost reasons and discussed as a confounder in §10. Neither Anthropic model was queried with extended thinking: our client code passes no thinking parameter on either the synchronous or the batch path. Snapshot identifiers are listed in Table 1.

### 4.4 Package Extraction

We extract package references in two ways. Install commands are matched by regular expressions: `pip install` anywhere in a response (first argument only) and `npm install` inside code blocks. Import statements are parsed from the code (Python's `ast` module; tree-sitter for JavaScript), and Python sub-imports are resolved to the top-level module. Scoped npm names (`@scope/name`) are kept whole. Other install forms (requirements files, `uv`, `poetry`, `yarn`, `pnpm`) are not extracted (§10). We follow Spracklen's extraction approach and inherit its known false-positive behavior; §4.6 removes the classes of false positives we could identify.

### 4.5 Validation Against the Live Registries

Earlier versions judged a name hallucinated if it was absent from the package-name lists shipped with the Spracklen artifact. Those lists predate our runs, so names registered in between were miscounted (§7). This version checks every candidate name against the live registries, with two methods per registry: for PyPI, the JSON API and the Simple API (PEP 691); for npm, the registry document and the deps.dev package API [24]. Every response is stored in an append-only log with its status code, a hash of its body and a UTC timestamp. A response of 200 counts as EXISTS and 404 as ABSENT; anything else (rate limiting, server errors, timeouts) counts as UNKNOWN. A name is finally ABSENT only if both methods return ABSENT in two rounds at least 12 hours apart (round 1: November 7, 2026, 11:53–13:14 UTC; round 2: November 8, 10:16–11:22 UTC); if the methods disagree, the name is UNKNOWN. Requests were limited to one per second per host. The two rounds made 14,821 requests without a single rate-limit or server error.

Hallucination is judged as of the moment of generation, not as of today. For each existing package we record its creation time (npm: `time.created`; PyPI: the earliest file upload) and compare it with the generation time of the name: the generation's own timestamp or, for batch requests, the window between batch submission and completion. A name whose package existed before generation is a false positive; a name whose package was created after generation is a hallucination that was registered later (§8); a creation time inside a batch window makes the name UNKNOWN. We also query OSV [23] for malicious-package records and keep npm security-holding placeholders, unpublished npm packages and PyPI projects without files as separate categories.

### 4.6 Exclusions

Before the live checks we remove four classes of names that are not package hallucinations; Table 7 counts each class.

1. *Extraction noise.* One noise filter, published with the code, removes syntactic prefixes (`package:`, `dart:`, `node:`), path aliases, template placeholders and path fragments.
2. *Standard-library and built-in modules:* the union of `sys.stdlib_module_names` for Python 3.10–3.13 and the built-in module lists of Node.js 18, 20 and 22. No hallucination candidate of earlier versions matched this rule, so standard-library names do not inflate our rates.
3. *Names that the registry would reject*, under PEP 508 for PyPI and npm's naming rules for new packages (for example, names containing whitespace or line breaks).
4. *Import names of real packages.* A Python name extracted only from import statements is not a hallucination if the pipreqs import-to-package mapping [22] maps it to a distribution that exists: the code is correct, and only the module name differs from the package name. A name that appears at least once as an install target stays a candidate, because installing it fetches whatever is registered under that exact name.

### 4.7 Rates and Statistical Analysis

We report two rates. The *mention-level* rate is the share of extracted package references that are hallucinated; it matches the unit of Spracklen et al. and of earlier versions. The *generation-level* rate is the share of all generations of a model that contain at least one hallucinated reference; this is the rate named in our preregistration. Both come with 95% Wilson confidence intervals [8, 11]. We compare models pairwise with a Pearson $\chi^2$ test on each 2×2 table [9], with Holm–Bonferroni correction [10] across the 10 pairs at $\alpha = 0.05$, and confirm each pair with a permutation test (10,000 resamples, seed 42). Overlap of the unique hallucinated names of two models is measured by Jaccard similarity.

### 4.8 Refusals

A generation is a refusal if the model declined or returned no usable code (an empty or "none" response). Refusal rates range from 1.27% (Claude Haiku 4.5) to 31.51% (GPT-5.4-mini at minimal reasoning effort). Refusals contain no package references: they do not enter the mention-level rate but lower the generation-level rate. We discuss the consequences in §10.

### 4.9 Preregistration

The revalidation and the registration analysis of §8 follow a plan committed to the project repository on November 7, 2026, before any live registry check (commit 686f648 in our private repository; the plan file's SHA-256, 18067fe7…c608, can be checked against the copy in the public artifact). The plan fixes the definitions above, the groups and the primary test of §8, and it was not edited afterwards. Every later change is recorded with its date and reason in a deviation log of 20 entries. Thirteen were written before the first live check (T0 = November 7, 2026, 11:44 UTC); among them are the import-name rule of §4.6 and the batch-window rule of §4.5. Six were written after T0 but before any group-level result was seen: the handling of npm placeholders and unpublished packages, a category for PyPI projects without files, the registry naming rules, the handling of removed malicious packages, and the use of arXiv submission times as exposure dates. One entry records a safety check made after the analysis (§8.5). The repository was local, so the commit times are self-attested; the plan, the log and the commit history are published with the artifact (§11.3).

## 5 Results

### 5.1 Headline Rates

Per-model mention-level rates range from 3.63% (Claude Haiku 4.5) to 5.02% (GPT-5.4-mini) (Figure 2a). All five lie between the lowest rate Spracklen et al. report for any model (3.59%, GPT-4 Turbo) and their commercial-model average (5.2%); the best 2026 model is practically level with the best 2024 model. Seven of the ten pairwise differences are significant after Holm–Bonferroni correction (Table 4), and the models fall into three tiers: Claude Haiku 4.5; Claude Sonnet 4.6; and a cluster of Gemini 2.5 Pro (4.77%), DeepSeek V3.2 (4.80%) and GPT-5.4-mini (5.02%) whose mutual differences are not significant. At the generation level (Figure 2b) rates range from 2.15% (GPT-5.4-mini) to 3.39% (Gemini 2.5 Pro), and 8 of 10 pairs differ significantly. The order changes mainly because GPT-5.4-mini refuses 31.51% of prompts, and refusals lower the generation-level rate.

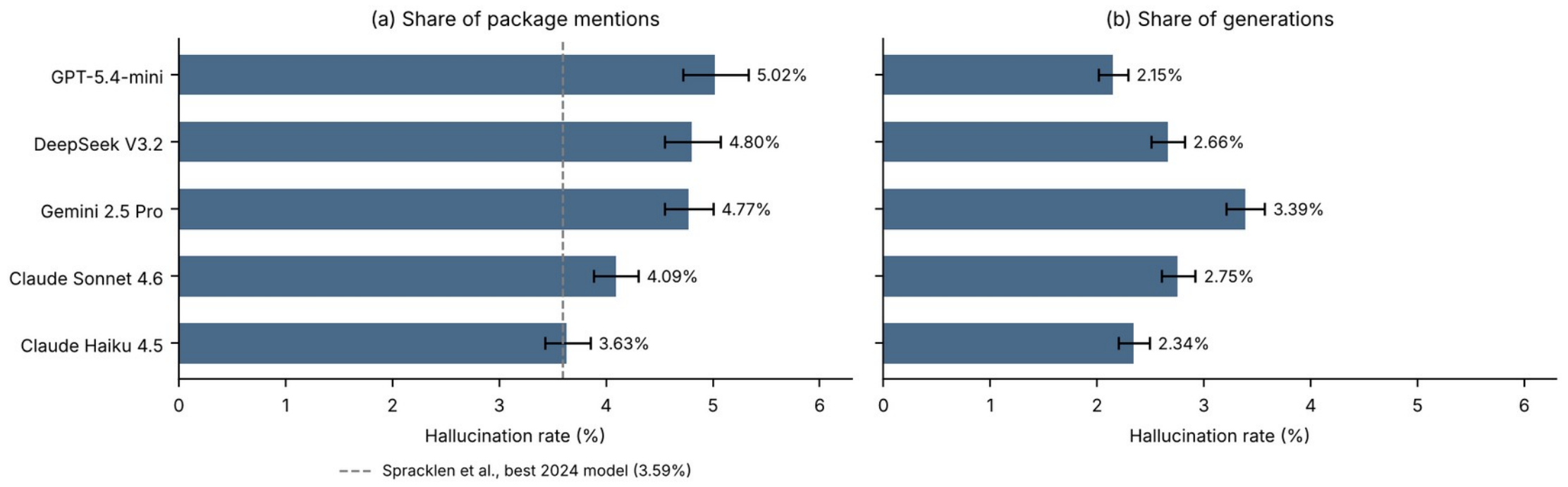

*Figure 2: Per-model hallucination rates with 95% Wilson confidence intervals: (a) share of package mentions, the unit of Spracklen et al., with their best 2024 model (GPT-4 Turbo, 3.59%) as a dashed line; (b) share of generations. Each model has 39,969 generations (Claude Sonnet 4.6: 39,964).*

## 5.2 Decomposition by Language and Dataset

Table 2 decomposes the rates by language. Python rates exceed JavaScript rates for every model at both levels: by 2.34 (Claude Haiku 4.5) to 3.47 percentage points (Claude Sonnet 4.6) per mention, and by 3.25 to 4.78 points per generation. This is the reverse of Spracklen's 2024 finding (§9.3). Table 3 decomposes the rates by prompt dataset: LLM-synthesized prompts produce higher rates than Stack Overflow prompts for every model; pooled over models, 4.79% (LLM_AT) and 5.44% (LLM_LY) against 2.28% (SO_AT) and 3.10% (SO_LY) of mentions.

*Table 2: Hallucination rates by language, in percent. Upper block: share of package mentions; lower block: share of generations. Differences carry 95% Newcombe intervals; per-cell Wilson intervals are in the artifact.*

| Model | Python | JavaScript | Δ (Py − JS), pp | Overall |
|---|---|---|---|---|
| Claude Haiku 4.5 | 4.38 | 2.04 | +2.34 [+1.93; +2.73] | 3.63 |
| Claude Sonnet 4.6 | 5.13 | 1.67 | +3.47 [+3.09; +3.84] | 4.09 |
| Gemini 2.5 Pro | 5.64 | 2.74 | +2.89 [+2.45; +3.32] | 4.77 |
| DeepSeek V3.2 | 5.51 | 2.93 | +2.59 [+2.06; +3.08] | 4.80 |
| GPT-5.4-mini | 5.95 | 2.67 | +3.28 [+2.68; +3.84] | 5.02 |
| *Per generation:* | | | | |
| Claude Haiku 4.5 | 4.15 | 0.69 | +3.46 [+3.16; +3.77] | 2.34 |
| Claude Sonnet 4.6 | 5.11 | 0.60 | +4.51 [+4.19; +4.85] | 2.75 |
| Gemini 2.5 Pro | 5.88 | 1.11 | +4.78 [+4.42; +5.15] | 3.39 |
| DeepSeek V3.2 | 4.68 | 0.81 | +3.88 [+3.56; +4.20] | 2.66 |
| GPT-5.4-mini | 3.85 | 0.59 | +3.25 [+2.97; +3.55] | 2.15 |

*Table 3: Hallucination rates by prompt dataset (share of package mentions, in percent). Generation-level rates and intervals are in the artifact.*

| Model | SO_AT | SO_LY | LLM_AT | LLM_LY |
|---|---|---|---|---|
| Claude Haiku 4.5 | 2.03 | 3.46 | 3.79 | 4.19 |
| Claude Sonnet 4.6 | 2.00 | 2.56 | 4.56 | 5.13 |
| Gemini 2.5 Pro | 2.67 | 3.04 | 5.37 | 6.15 |
| DeepSeek V3.2 | 2.68 | 4.15 | 4.97 | 5.70 |
| GPT-5.4-mini | 1.52 | 1.82 | 5.37 | 6.28 |
| All models | 2.28 | 3.10 | 4.79 | 5.44 |

### 5.3 Pairwise Model Differences

Table 4 reports all ten pairwise comparisons. The largest difference is between Claude Haiku 4.5 and GPT-5.4-mini ($\Delta$ = 1.38 pp, $\chi^2$ = 56.46, Holm-corrected $p < 10^{-4}$). Every pair involving Claude Haiku 4.5 or Claude Sonnet 4.6 is significant; the three pairs within the Gemini–DeepSeek–GPT cluster are not ($\Delta \leq 0.25$ pp). The permutation tests agree with the $\chi^2$ tests on every pair.

*Table 4: Pairwise differences of mention-level rates (percentage points), with $\chi^2$ statistics and Holm–Bonferroni-corrected p-values across the 10 pairs ($\alpha$ = 0.05). *** $p < 0.001$; * $p < 0.05$; ns, not significant. The last column gives the result at the generation level.*

| Pair | $\Delta$, pp | $\chi^2$ | p (Holm) | Sig. | Per generation |
|---|---|---|---|---|---|
| Haiku 4.5 vs GPT-5.4-mini | 1.38 | 56.46 | $< 10^{-4}$ | *** | ns ($\Delta$ 0.20) |
| Haiku 4.5 vs DeepSeek V3.2 | 1.17 | 47.47 | $< 10^{-4}$ | *** | sig. ($\Delta$ 0.32) |
| Haiku 4.5 vs Gemini 2.5 Pro | 1.13 | 50.58 | $< 10^{-4}$ | *** | sig. ($\Delta$ 1.05) |
| Sonnet 4.6 vs GPT-5.4-mini | 0.93 | 25.31 | $< 10^{-4}$ | *** | sig. ($\Delta$ 0.61) |
| Sonnet 4.6 vs DeepSeek V3.2 | 0.72 | 17.88 | 0.0001 | *** | ns ($\Delta$ 0.09) |
| Sonnet 4.6 vs Gemini 2.5 Pro | 0.68 | 18.63 | $< 10^{-4}$ | *** | sig. ($\Delta$ 0.64) |
| Haiku 4.5 vs Sonnet 4.6 | 0.45 | 8.76 | 0.012 | * | sig. ($\Delta$ 0.41) |
| Gemini 2.5 Pro vs GPT-5.4-mini | 0.25 | 1.61 | 0.613 | ns | sig. ($\Delta$ 1.24) |
| DeepSeek V3.2 vs GPT-5.4-mini | 0.22 | 1.06 | 0.613 | ns | sig. ($\Delta$ 0.51) |
| Gemini 2.5 Pro vs DeepSeek V3.2 | 0.03 | 0.03 | 0.862 | ns | sig. ($\Delta$ 0.73) |

### 5.4 Inter-Model Overlap of Hallucinated Names

Figure 3 reports the Jaccard overlap of the unique hallucinated names of each pair of models. The mean is 0.219 (range 0.173–0.342). The highest overlap is between DeepSeek V3.2 and GPT-5.4-mini (J = 0.342; 344 shared names out of 1,005), well above every other pair, which suggests either shared training-data origins or convergent naming patterns; we cannot distinguish the two. The lowest overlaps involve Claude Haiku 4.5: with Gemini 2.5 Pro (J = 0.173) and with Claude Sonnet 4.6 (J = 0.176).

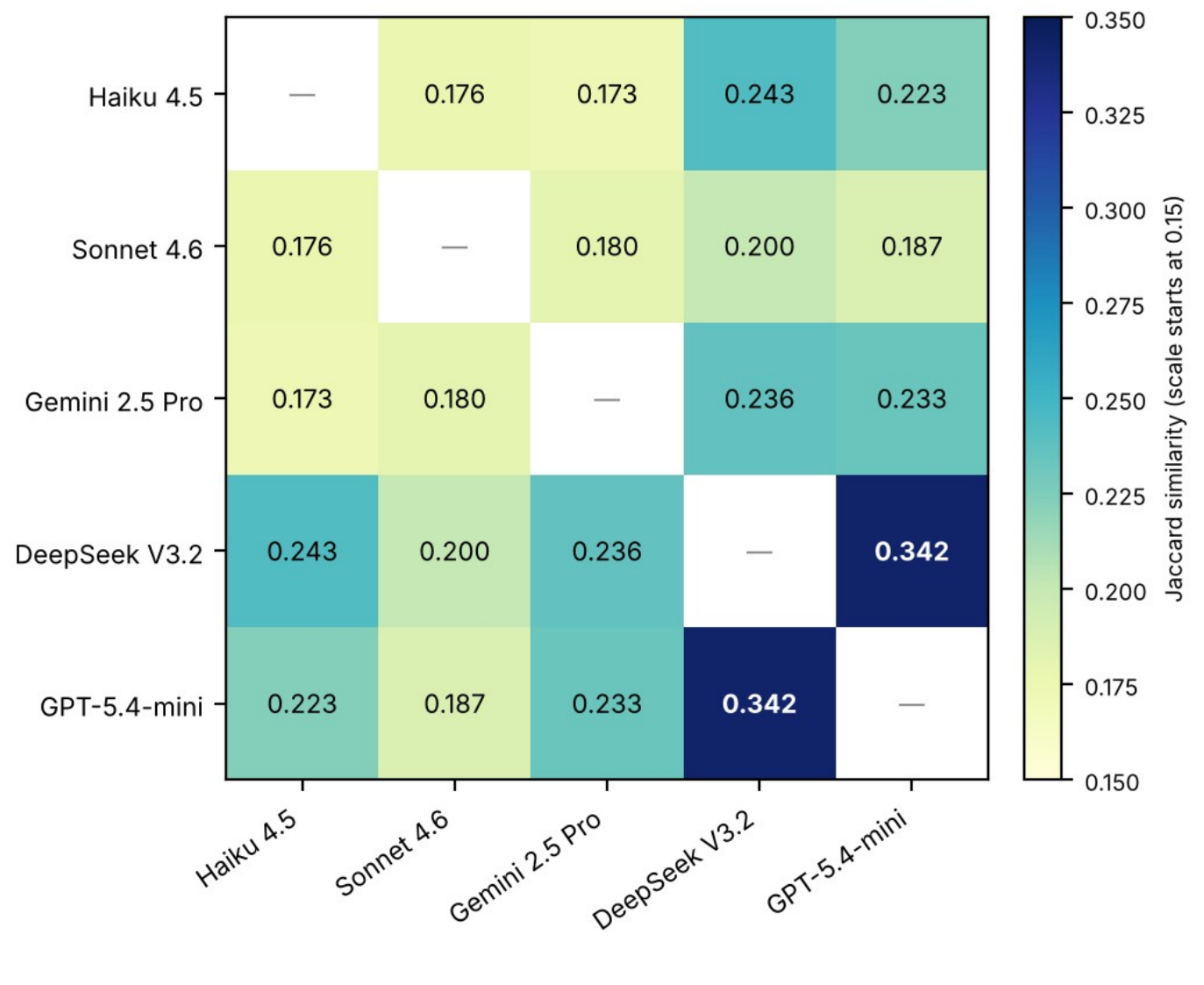

*Figure 3: Pairwise Jaccard similarity between the sets of unique hallucinated package names of two models. The color scale starts at 0.15 so that values between 0.17 and 0.34 can be told apart. The diagonal is omitted.*

### 5.5 The Haiku-below-Sonnet Inversion

Claude Haiku 4.5 hallucinates less than Claude Sonnet 4.6 at both levels (3.63% vs 4.09% of mentions, Holm-corrected $p = 0.012$; 2.34% vs 2.75% of generations), contrary to Spracklen's observation that smaller models within a family tend to hallucinate more. Earlier versions explained the inversion in part by Haiku 4.5 having been queried with extended thinking. We withdraw that explanation: our client code passes no thinking parameter to either model (§4.3). Differences in post-training between the two models remain a possible cause that we cannot test. The low overlap between the two models ($J = 0.176$) suggests that they make different errors rather than one making a subset of the other's; we have no mechanism for the inversion.

### 5.6 Composition of the Universal Set

Table 5 breaks the 104 universal names (§6) down by category. Four npm names lie in scopes owned by an organization, such as `@ember/*`; no outsider can register a name in such a scope. Fifty-four names are one edit (Levenshtein distance 1 after normalization) away from a registered package, and they account for most of the set's hallucinated mentions (1,571 of 2,699). We use distance 1 rather than the distance 2 of earlier versions: with about 4.2 million npm and 857,000 PyPI names registered, distance 2 matches 81 of the 104 names and says little. This version prints no other universal name. Earlier versions printed 18 names from their universal list, 13 of which remain in the current set; those names are therefore public, and we do not repeat or extend the list.

*Table 5: The universal-hallucination set by category. Categories are mutually exclusive and assigned in row order. "One edit away" means Levenshtein distance 1, after normalization, to a name in the registries' package lists of July 23, 2026. "Mentions" counts hallucinated mentions across the five models.*

| Category | PyPI names | PyPI mentions | npm names | npm mentions | Names | Mentions |
|---|---|---|---|---|---|---|
| npm scope owned by an organization | 0 | 0 | 4 | 148 | 4 | 148 |
| One edit away from a registered package | 49 | 1,546 | 5 | 25 | 54 | 1,571 |
| Other | 39 | 944 | 7 | 36 | 46 | 980 |
| **Total** | **88** | **2,490** | **16** | **209** | **104** | **2,699** |

## 6 Universal Hallucinations as a Model-Agnostic Attack Vector

The *universal-hallucination set* contains the names that all five models invent. We define it as the intersection of the per-model sets of names that were hallucinations at generation time, after the exclusions of §4.6. It contains 104 names (88 PyPI, 16 npm), which account for 2,699 hallucinated mentions across the five models.

In April 2026 we disclosed the earlier, larger list to PyPI Security and to Socket.dev and asked which names an outsider could actually register (§11.1). Socket's review of the 18 npm names found six non-exploitable: the four `@ember/*` names, which are Ember.js virtual modules in a scope owned by the Ember organization; one already-published package; and one npm security-holding placeholder. Our revalidation independently classifies the last two as false positives, so the 16 npm names of the current set are Socket's 12 registrable candidates plus the four `@ember/*` names. On the PyPI side, PyPI Security reported that its prohibited-name list and normalization rules left 41 of the disclosed names registrable, without listing which. Because the PyPI part of the set has changed (§7), we do not re-estimate PyPI registrability; we have asked PyPI Security for the list.

Universal names carry a risk that no single-model rate shows. By studying all five models, an attacker finds names that reach users of several providers: a single registration of a registrable universal name would be installed regardless of which of the five models the victim uses.

We suggest two non-exclusive mechanisms by which a name becomes universal. First, shared training data: tutorials, posts and documentation that refer to a package by a module or informal name instead of its distribution name become a stable signal across providers, which scrape similar public corpora. Second, overgeneralized naming conventions: scoped npm names such as `@ember/service` follow a convention that models extend to install targets, although Ember ships those modules only inside its main package. Both mechanisms lead different providers to the same wrong name, so post-training at any one provider does not remove it.

The DeepSeek V3.2–GPT-5.4-mini Jaccard peak (J = 0.342, §5.4) fits this picture: the shared errors are a small fraction of any one model's set overall, but between certain pairs they reach one third. We make no inference about the direction of any data-sharing relationship.

A single-model measurement therefore understates the slopsquatting attack surface: it cannot tell a name that one model hallucinates from a name that every comparable model hallucinates. We recommend reporting cross-cohort intersections as a standard quantity.

## 7 What Changed Since Earlier Versions

Versions 1–3 differ from this version in three ways: names were validated against static 2025 package lists rather than the live registries; the import names of real packages were counted as package names; and some numbers were not reproducible from the stored data. Table 6 lists every changed headline number with its cause, and Table 7 counts the names removed at each step.

Rates fell by about a fifth for every model. For Claude Haiku 4.5, hallucinated mentions fell from 1,394 to 1,098: 189 were import names of real packages, 51 belonged to packages that already existed when the name was generated, 44 were names no registry accepts, 11 are UNKNOWN, and 1 was a spelling variant of an existing package. Import names of real packages are the largest correction for every model (144–207 mentions per model).

The universal set fell from the 127 names reported earlier to 104. The earlier figure itself did not reproduce: the stored data give 121 PyPI names, not 109, so the earlier list had 139 names. Of the 121 PyPI names, 88 remain; 26 came only from correct import statements, 3 belonged to packages that already existed, 1 was an extraction artifact, and 3 are UNKNOWN (one PyPI project without files and two removed malicious packages whose creation dates are unknown). Of the 18 npm names, 16 remain; the other 2 belonged to packages that already existed.

Three smaller corrections apply. The corpus has 199,840 generations, not 199,845: five Claude Sonnet 4.6 generations are missing from the stored data, and the earlier per-provider counts for Google (−1) and DeepSeek (+6) do not match the stored data. The exact rates of earlier versions (4.62–6.10%) are not reproducible even with the earlier method, which gives 4.61–6.08% on the stored data. Refusal rates change slightly under one uniform definition (Table 1).

The qualitative conclusions are unchanged: Python exceeds JavaScript for every model, synthesized prompts exceed Stack Overflow prompts, Claude Haiku 4.5 has the lowest mention-level rate, and a large universal set exists.

*Table 6: Changes from versions 1–3 to version 4. Version 4 rates are given as share of mentions / share of generations.*

| Quantity | Versions 1–3 | Version 4 | Main reason |
|---|---|---|---|
| Generations | 199,845 | 199,840 | 5 Claude Sonnet 4.6 generations missing; earlier total was prompts × 5 |
| Claude Haiku 4.5 | 4.62% | 3.63% / 2.34% | 1,394 → 1,098 mentions: −189 import names, −51 existed, −44 invalid, −11 unknown, −1 variant |

| Quantity | Versions 1–3 | Version 4 | Main reason |
|---|---|---|---|
| Claude Sonnet 4.6 | 5.41% | 4.09% / 2.75% | 1,866 → 1,407: −202 import names, −169 invalid, −63 existed, −23 unknown, −2 other |
| Gemini 2.5 Pro | 5.80% | 4.77% / 3.39% | 1,974 → 1,625: −207 import names, −77 existed, −49 invalid, −14 unknown, −2 other |
| DeepSeek V3.2 | 5.89% | 4.80% / 2.66% | 1,531 → 1,251: −164 import names, −63 existed, −36 invalid, −15 unknown, −2 other |
| GPT-5.4-mini | 6.10% | 5.02% / 2.15% | 1,190 → 982: −144 import names, −45 existed, −12 unknown, −7 invalid |
| Range | 4.62–6.10% | 3.63–5.02% / 2.15–3.39% | earlier exact values not reproducible (earlier method: 4.61–6.08%) |
| Significant pairs | 5 of 10 | 7 of 10 / 8 of 10 | recomputed |
| Jaccard | 0.222 (0.179–0.343) | 0.219 (0.173–0.342) | sets after revalidation |
| Universal set | 127 (109 + 18) | 104 (88 + 16) | 109 not reproducible (data give 121); import names, existing packages, unknown |
| Registrable subset | 53 | not re-estimated | registrability is decided by the registries (§6) |
| Refusals (Haiku / Sonnet / Gemini / DeepSeek / GPT) | 1.27 / 18.79 / 8.14 / 2.08 / 32.14% | 1.27 / 18.77 / 8.11 / 1.97 / 31.51% | one uniform refusal definition |

*Table 7: From the candidate names of versions 1–3 to the hallucinated names of version 4 (unique names, all models).*

| Step | PyPI | npm |
|---|---|---|
| Hallucination candidates in versions 1–3 | 2,299 | 1,146 |
| Removed: extraction noise | −313 | −383 |
| Removed: standard-library or built-in module | 0 | 0 |
| Removed: invalid under the registry’s naming rules | −81 | −150 |
| Removed: import name of a real package | −54 | — |
| Checked against the live registries | 1,851 | 613 |
| … hallucination at generation (absent today) | 1,724 | 533 |
| … hallucination at generation, registered later | 22 | 3 |
| … false positive (package existed before generation) | 84 | 77 |
| … UNKNOWN | 21 | 0 |
| **Unique hallucinated names (v4)** | **1,746** | **536** |

# 8 Do Hallucinated Names Get Registered?

## 8.1 A Natural Experiment

Our list of universal names became public by mistake on April 30, 2026, one day after we disclosed it to PyPI Security and Socket (§11.2). It stayed public in our repository and in an open archive until November 7, 2026, and a third-party fork still holds a copy. Ten names printed in version 1 of this paper were public from May 16, 2026, and further names printed in version 2 from June 11, 2026. All other hallucinated names were never published. This gives a natural experiment. If publishing hallucinated names attracts registrations, published names should be registered more often than comparable unpublished ones.

We fixed the design in the preregistration (§4.9) before checking any registry. Group A is the universal set (104 names), split into A1 (13 names printed in the paper) and A2 (91 names that were only in the list). Group B contains all other hallucinated names (2,178); its subset B4, the main comparison group, contains the names in B that exactly four of the five models hallucinate (128). The outcome is registration between the start of exposure and T0 (November 7, 2026). Exposure starts at the arXiv submission of the first version that printed the name (A1) or on April 30, 2026 (A2 and B). The primary test compares A with B4 within common strata of name frequency and name length, with Fisher's exact test (two-sided).

### 8.2 Results

No effect of publication is visible at this precision (Table 8). In the primary comparison, 0 of 87 names in A and 2 of 101 names in B4 were registered: a difference of −2.0 percentage points (95% CI −6.9 to +2.5; $p = 0.50$). The interval excludes an increase above 2.5 points. Stratified tests (Cochran–Mantel–Haenszel, $p = 0.59$ by ecosystem and $p = 0.72$ by stratum) and all preregistered sensitivity analyses lead to the same conclusion. With these group sizes, the study could detect only a rise from a 1% baseline to about 10.5%.

Overall, 25 of the 2,282 names that were hallucinations at generation were registered within about six months: 1 of 104 published names (1.0%) and 24 of 2,178 unpublished names (1.1%). The median time from the start of exposure to registration was 89 days in B (range 13–153); the one registered name in A was registered after 47 days.

*Table 8: Registrations of hallucinated names between the start of exposure and T0. Differences carry 95% Newcombe intervals. Group sizes: A 104 (A1 13, A2 91), B 2,178, B4 128; the primary test uses the 11 strata that contain names of both groups. Minimum detectable registration share in A (power 0.8): 10.5%, 12.5% and 18.5% at baseline shares of 1%, 2% and 5%.*

| Comparison | First group | Second group | Difference, pp | Fisher p |
|---|---|---|---|---|
| **Primary: A vs B4 (common strata)** | 0/87 (0.0%) | 2/101 (2.0%) | −2.0 [−6.9; +2.5] | 0.50 |
| PyPI only | 0/76 | 2/97 | −2.1 [−7.2; +3.0] | 0.51 |
| npm only | 0/11 | 0/4 | 0 [−49.0; +25.9] | 1.00 |
| A vs all of B (descriptive) | 1/104 (1.0%) | 24/2,178 (1.1%) | −0.1 [−1.1; +4.2] | 1.00 |
| A1 vs A2 (descriptive) | 1/13 (7.7%) | 0/91 (0.0%) | +7.7 [+0.2; +33.3] | 0.13 |

### 8.3 What Was Registered

We classify registered packages from registry metadata only. Of the 24 registrations in B, 22 are published packages with files whose content we did not inspect; one is an npm security-holding placeholder whose package carries a malicious-package record published after our runs, so it was registered with malware and removed by npm; and one is an empty placeholder whose publisher describes it as a name reservation. The single registration in A, a name printed in version 2 of this paper, is a package with one version; a static check (§8.5) found it empty. We do not infer intent from registrations: researchers and defenders also reserve names.

For methods that use the registry as an oracle [21], these results mean that the precondition for oracle poisoning does occur, but rarely: about 1% of hallucinated names were registered within six months, and at least one registration carried malware.

### 8.4 Names with a Malware History

Packages registered before our runs matter as well. Thirteen npm names that the models suggest were, at generation time, npm security-holding placeholders: packages that existed before our runs, carry malicious-package records in OSV, and were replaced by npm. On PyPI, ten names that the models suggest have OSV malicious-package records whose identifiers date from before 2026; the packages are gone. In total, at least 23 package names that the models suggest had hosted malware that the registries later removed. Fifteen further npm names belong to packages that were unpublished, 14 of them before our runs; two of these also have malicious-package records. Such names are natural slopsquatting targets, because the models still suggest them after the registries have removed the malware.

### 8.5 Safety Check

Because the one registered name in group A was printed in this paper, we checked it statically after the analysis (deviation D19): we downloaded and listed its archive without installing or running it. The archive contains only a package manifest, with no scripts, no dependencies and no code.

## 9 Discussion

### 9.1 Why Has the Range Compressed?

Our five models span 1.39 percentage points (3.63–5.02% of mentions). In 2024 the class averages alone were 5.2% for commercial and 21.7% for open-source models. We propose three contributing causes. First, the gap between open and closed models that drove the upper end of the 2024 range has closed at the frontier: DeepSeek V3.2 (4.80%) sits inside the commercial cluster. Second, training-data curation for package references may be near saturation: the lower end of our range (3.63%) matches Spracklen's best 2024 model (3.59%) almost exactly. Third, post-training recipes have converged across providers, reducing behavioral variance.

### 9.2 Why Does the Threat Persist?

A rate of 3.6–5.0% keeps slopsquatting economically viable. Even the lowest-rate model produced 1,098 hallucinated mentions over 39,969 prompts, covering 605 unique names. A single popular hallucinated name, reproduced under similar prompts, yields a steady stream of install attempts once a malicious package is registered, and registration is free. Section 8 shows that hallucinated names are in fact registered, at least one with malware, and that names which once hosted malware are still suggested.

### 9.3 Python Worse Than JavaScript: An Inversion

Spracklen reported JavaScript as the noisier ecosystem, attributing the effect to npm's larger and more fragmented namespace. We find the opposite for every 2026 model, with Python exceeding JavaScript by 2.34–3.47 percentage points of mentions. The asymmetry survives the removal of import names of real packages (§4.6), so it is not an artifact of module-name confusion. One plausible mechanism is that 2026 models have absorbed npm's flat naming convention more thoroughly than Python's mix of single-token, snake_case, dotted and dash-separated names, and of distribution names that differ from module names. With more ways to combine name parts, a model more often produces a plausible but wrong name.

### 9.4 Synthesized Prompts Surface More Hallucinations

LLM-synthesized prompts produce higher rates than curated Stack Overflow prompts for every model (Table 3). A cautious reading is that synthetic prompts drift away from the training distribution and reveal failure modes that organic developer questions do not. Rates measured on curated questions alone are therefore lower bounds on what deployment will see, and an attacker gets more hallucinated names from synthesized prompts than from Stack Overflow questions.

### 9.5 Dependence on Model Selection

We sampled five frontier models, and the compression is conditional on that selection. Models from the next tier down, such as smaller Qwen, Mistral or older Llama checkpoints still in production, may still show 2024-like rates. The finding applies to the frontier; we do not claim that the attack surface has narrowed at every capability tier.

## 10 Limitations and Threats to Validity

*Scope.* We evaluate five frontier models; smaller open models, fine-tuned variants and quantized local checkpoints are out of scope, as are agentic configurations in which retrieval could eliminate hallucinations.

*Timing.* Generations were collected on April 22–28, 2026, and registries were checked on November 7–8, 2026. Models change opaquely, and our rates describe one point in time.

*Leakage.* The Spracklen prompt corpus, released in January 2025, may be part of 2026 pretraining data, which would bias rates downward. We did not run a held-out set of new prompts.

*GPT-5.4-mini.* It was run at minimal reasoning effort and refused 31.51% of prompts. Refusals do not enter the mention-level rate but lower the generation-level rate, which is why the two levels rank this model differently. The compliant responses may not be a random subset of prompts, and we did not sweep reasoning effort.

*DeepSeek.* DeepSeek does not pin model versions in its public API. The deepseek-chat alias was nominally V3.2 during our run, but same-window response metadata returned the identifier deepseek-v4-flash; the underlying model may have changed during or shortly after the run [7].

*Run configuration.* The generation code in the repository is unchanged since April 28, 2026, but the repository alone cannot prove that this exact code produced every generation.

*Extraction.* Install commands are extracted only from `pip install` (first argument) and from `npm install` inside code blocks; other forms are missed, so rates may be underestimated.

*Import names.* The pipreqs mapping covers about 1,157 common import names. Rarer mismatches between module and package names are still counted as hallucinations, mostly among names that few models produce.

*Creation dates.* The PyPI creation date is the earliest visible upload and can be late if files were deleted (no such case was flagged). Removed malicious packages have no creation date and are UNKNOWN; for them we use the year in the malicious-package record identifier as an approximate date.

*Registrability.* It is known only from the review of the earlier list by PyPI Security and Socket in April–May 2026 and was not re-estimated.

*Natural experiment.* It is small and can detect only large effects; exposure intensity is only partly known (20 downloads of the open archive, one public fork); and registrations by researchers or defenders cannot be told apart from attacks without inspecting content.

*Preregistration.* The plan and the deviation log were committed to a local repository, so their timestamps are self-attested.

## 11 Ethics and Open Science

### 11.1 Ethical Considerations

This study produces lists of hallucinated package names that, if registered, could be weaponized into a supply-chain attack. We followed a stakeholder-based analysis [12] covering developers who receive hallucinated suggestions, the LLM providers, the PyPI and npm maintainers, and downstream users of compromised software. To minimize harm we (a) registered no names on any registry; (b) disclosed the enumerated universal list to PyPI Security (security@pypi.org) and to Socket.dev (ticket #2893) on April 29, 2026, intending a private window before any publication (npm's security mailbox no longer accepted external reports, and its support portal redirects registry-level reports to a bug-bounty program unsuited to academic disclosure); (c) plan to share model-specific results with each provider; (d) in this version print no hallucinated name that an outsider could register, whereas earlier versions printed a top-10 table and a few examples, which are therefore already public; and (e) installed and ran no third-party package, reading registry metadata only, apart from the static check of §8.5. PyPI Security and Socket reviewed the list (§6). The study used no human subjects, so IRB review did not apply. The author has no conflicts of interest with the studied providers.

### 11.2 Accidental Exposure of the List

Our coordinated disclosure did not go as planned. On April 30, 2026, one day after the disclosure, the full universal list was committed by mistake to a `disclosure/` directory of our public artifact repository and archived in an open Zenodo version, which was downloaded 20 times. We found the mistake on July 23 and informed Socket; the directory was removed from the current tree in August but remained in the repository history and in the archive. On November 7, 2026, we removed it from the full history, restricted the archived files, informed PyPI Security of the exposure and Socket that it had lasted longer than we had reported, and asked GitHub to remove cached copies. A third-party public fork created on July 26 still contains the list. We therefore treat the list as public since April 30, 2026. Section 8 uses this exposure as a natural experiment, but the list was not kept public for that purpose.

### 11.3 Open Science

The replication code is at https://github.com/churik5/slopsquatting-replication-2026 (MIT license); its history was rewritten on November 7, 2026, so commit references from earlier versions no longer resolve. The Zenodo concept DOI 10.5281/zenodo.19859120 resolves to the latest archived version; the files of version v0.2 are restricted because they contain the list. The revalidation code, the preregistration with its deviation log and commit history, and all aggregate results of this version are published at https://github.com/churik5/slopsquatting-revalidation-2026. No file in the artifact contains a registrable hallucinated name. The SQLite database of raw model responses is available to verified researchers on request via churik0509@gmail.com, mirroring the policy of [1]; earlier versions stated that it had been deposited in an access-controlled Zenodo record, which was not the case.

## 12 Conclusion

We replicated Spracklen et al. on five frontier code-generating LLMs released between October 2025 and March 2026 and revalidated every name against the live registries. The range has shrunk: the five models hallucinate 3.6–5.0% of package references, at the level of 2024’s best commercial models. Still, 104 names are hallucinated by all five models, one registration of a registrable name can reach users of all four providers, and about 1% of hallucinated names were registered within six months (at least one with malware), whether or not anyone had published them. At least 23 names the models suggest had already hosted malware. Single-model rate measurements understate the attack surface; cross-cohort intersections and registration monitoring should be standard parts of future measurements.

## Acknowledgments

We thank the authors of Spracklen et al. [1] for releasing their prompt datasets, package lists and validation code, without which this replication would not have been possible. We thank Mike Fiedler of PyPI Security and the Socket.dev security team for their review of the universal-hallucination set. The author funded this study from personal accounts, apart from a one-time Google Cloud free credit grant for Gemini 2.5 Pro inference. No institutional, governmental or industry funding was received. The author used AI assistants (Anthropic’s Claude) to write analysis code, to draft and edit the text of this version, and to check its consistency; the author reviewed all content and is responsible for it.

## Appendix A: Generations, API Calls and Cost

Our pipeline stored the token counts and an estimated cost for every API call (tokens times the list price, with batch discounts where batches were used). The estimates sum to $1,160.29. This includes $17.55 for an early partial GPT-5.4-mini run on April 22–24 (15,517 calls) whose generations were discarded. Earlier versions reported a total of $860.90. The two figures agree for OpenAI and DeepSeek; the stored data do not reproduce the earlier Anthropic and Google figures (Table A.1).

*Table A.1: Generations, API calls and API cost per provider, from the stored data and as reported in versions 1–3. Snapshot identifiers are listed in Table 1. The OpenAI calls include the discarded partial run.*

| Provider | Generations (stored) | Versions 1–3 | API calls | Estimated cost (USD) | Versions 1–3 (USD) |
|---|---|---|---|---|---|
| Anthropic (Sonnet 4.6 + Haiku 4.5) | 79,933 | 79,933 | 79,933 | 629.00 | 375.26 |
| OpenAI (GPT-5.4-mini) | 39,969 | 39,969 | 55,486 | 93.54 | 93.54 |
| Google (Gemini 2.5 Pro) | 39,969 | 39,968 | 39,969 | 426.64 | 380.99 |
| DeepSeek (V3.2) | 39,969 | 39,975 | 39,969 | 11.11 | 11.11 |
| **Total** | **199,840** | **199,845** | **215,357** | **1,160.29** | **860.90** |

## Appendix B: Prompt Examples

Earlier versions showed four illustrative examples. We replace them with real examples from the stored corpus, each with its prompt identifier so that it can be checked. To avoid printing names that an outsider could register, we show only generations in which every hallucinated name lies in an npm scope that already belongs to someone else, so the examples are not a random sample; all four are JavaScript, because PyPI has no scoped names. For each dataset we took the first eligible generation in seed-42 order, except that for SO_LY we skipped a generation whose invented name was an obvious placeholder, and for LLM_LY one that repeated the name of the LLM_AT example. Prompts are paraphrased.

**Example 1 (SO_AT; Gemini 2.5 Pro; prompt `javascript:SO_AT:4802`).** The prompt asks for a tool that draws a diagram of an existing relational database schema. The response contains:

```
// First, you need to convert your MySQL schema to DBML.
// You can use a tool like 'sql2dbml'
(https://www.npmjs.com/package/@dbml/sql-to-dbml).
```

The package `@dbml/sql-to-dbml` does not exist on npm (both check rounds). Other packages are already published under the `@dbml` scope, so no one else can publish this name.

**Example 2 (SO_LY; Gemini 2.5 Pro; prompt `javascript:SO_LY:3846`).** The prompt asks how to collect code coverage from end-to-end browser tests in a monorepo and report it together with unit-test coverage. The response contains:

```
{
  loader: '@js-bits/coverage-istanbul-loader',
  options: {
```

The package `@js-bits/coverage-istanbul-loader` does not exist on npm; the `@js-bits` scope already holds other packages.

**Example 3 (LLM_AT; Claude Haiku 4.5; prompt `javascript:LLM_AT:2292`).** The prompt asks for code that sets up an observability SDK with export of traces and metrics and error logging. The response contains:

```
const { NodeSDK } = require('@opentelemetry/auto');
const { getNodeAutoInstrumentations } =
require('@opentelemetry/auto-instrumentations-node');
```

The first package does not exist on npm; the second does. The `@opentelemetry` scope already holds 170 packages.

**Example 4 (LLM_LY; Claude Haiku 4.5; prompt `javascript:LLM_LY:2227`).** The prompt asks for a base class for a server-side component that instrumentation modules can extend, with shared setup and teardown for tracing and metrics. The response contains:

```
const { MeterProvider, PeriodicExportingMetricReader } =
require('@opentelemetry/sdk-metrics');
const { JaegerExporter } = require('@opentelemetry/exporter-trace-
jaeger');
const { OTLPMetricExporter } = require('@opentelemetry/exporter-
metrics-otlp-http');
```

The second package does not exist on npm; the first and third do.